\documentclass[times]{speauth}

\usepackage{fix-cm}
\usepackage{moreverb}
\usepackage{graphicx}
\usepackage{subfigure}
\usepackage{multirow}
\usepackage[perpage]{footmisc}
\usepackage{booktabs}
\usepackage{multirow}
\usepackage[colorlinks,bookmarksopen,bookmarksnumbered,citecolor=red,urlcolor=red]{hyperref}
\usepackage{verbatim}

\newcommand\fw{APLCache}

\newcommand\BibTeX{{\rmfamily B\kern-.05em \textsc{i\kern-.025em b}\kern-.08em
T\kern-.1667em\lower.7ex\hbox{E}\kern-.125emX}}

\def\volumeyear{2010}

\begin{document}

\runningheads{J. Mertz and I. Nunes}{Automation of Application-level Caching in a Seamless Way}

\title{Automation of Application-level Caching in a Seamless Way}

\author{Jhonny Mertz\affil{1}\corrauth and Ingrid Nunes\affil{1,2}}

\address{\affilnum{1}Universidade Federal do Rio Grande do Sul (UFRGS), Porto Alegre, Brazil
\break
\affilnum{2}TU Dortmund, Dortmund, Germany}

\corraddr{Universidade Federal do Rio Grande do Sul (UFRGS), 9500, Bento Gonçalves Avenue, Porto Alegre, Brazil. E-mail: jmamertz@inf.ufrgs.br}


\begin{abstract}

Meeting performance and scalability requirements while delivering services is a critical issue in web applications. Recently, latency and cost of Internet-based services are encouraging the use of \emph{application-level caching} to continue satisfying users' demands and improve the scalability and availability of origin servers. Application-level caching, in which developers manually control cached content, has been adopted when traditional forms of caching are insufficient to meet such requirements. Despite its popularity, this level of caching is typically addressed in an \emph{ad-hoc} way, given that it depends on specific details of the application. Furthermore, it forces application developers to reason about a crosscutting concern, which is unrelated to the application business logic. As a result, application-level caching is a time-consuming and error-prone task, becoming a common source of bugs. Among all the issues involved with application-level caching, the decision of what should be cached must frequently be adjusted to cope with the application evolution and usage, making it a challenging task. In this paper, we introduce an automated caching approach to automatically identify application-level cache content at runtime, by monitoring system execution and adaptively managing caching decisions. Our approach is implemented as a framework that can be seamlessly integrated into new and existing web applications. In addition to the reduction of the effort required from developers to develop a caching solution, an empirical evaluation showed that our approach significantly speedups and improves hit ratios, with improvements ranging from 2.78\% to 17.18\%.

\end{abstract}


\maketitle


\section{Introduction}


With the increasing popularity of web applications and software systems distributed on top of the web, it is crucial to improve their performance and scalability due to a large number of users. When traditional caching solutions are unable to meet performance requirements, \emph{application-level caching} is adopted to store content at a granularity that is possibly best suited to the application, thus allowing developers to separate common from customized content at a fine-grained level. It thus has become a popular technique to reduce the workload on content providers, which can thus decrease the user perceived latency. However, deciding the right content to cache and the best moment of caching is a challenging task, given that it depends on extensive knowledge of application specificities to be done efficiently. Otherwise, caching may not improve application performance or even may lead to a performance decay~\cite{Ports2010}. Furthermore, developers must continuously inspect application performance and revise caching design choices, due to changing workload characteristics and access patterns. Consequently, initial caching decisions may become obsolete over time~\cite{Chen2016,Radhakrishnan2004}.

Because application-level caching is essentially a manual task, its design and implementation are time-consuming and error-prone. Moreover, because its implementation is often interleaved with the business logic, it decreases code understanding, thus being a common source of bugs. Existing research addresses these limitations through static and dynamic analyses that identify caching opportunities, such as web pages~\cite{Negrao2015}, and database queries and objects~\cite{Chen2016}. By focusing on particular bottlenecks, proposed approaches help developers while addressing caching problems, but complex logic and personalized web content remain unaddressed. Recent work focuses on identifying cacheable methods that repeatedly perform the same computation~\cite{DellaToffola2015}, but it is limited to the suggestion of performance fixes, and developers should review suggestions and manually refactor the code, inserting cache logic into the application.

We thus, in this paper, propose a novel seamless and automated approach that chooses and manages cacheable content according to observations made by monitoring web applications at runtime, adding automatic and adaptive caching that leads to statistically significant speedups and hit-ratios. The automatically selected cache configuration reflects the monitored application workload, thus being sensitive to the application changing dynamics and self-optimizing caching decisions. As opposed to traditional caching approaches, our proposal focuses on caching \emph{method-level content}. Moreover, we use \emph{application-specific information} to make caching decisions, such as the user session, which are information that developers take into consideration while developing a caching solution. Thus, our approach can potentially reduce the reasoning and effort required from developers. In addition to the reduction of the effort required from developers to develop a caching solution, our approach is implemented as a framework, named \fw{}, which seamlessly integrates the proposed solution to web applications. Consequently, our approach and framework can prevent code tangling and raise the abstraction level of caching as well as detach caching concerns from the application.

We evaluated our approach empirically with three open-source web applications, which have different domains and sizes. Obtained results indicate that our approach can identify adequate caching opportunities by improving application throughput by factors between 2.78\%--17.18\%. Our approach can thus support developers by providing an automated approach to address issues related to the development of an application-level caching solution. As opposed to related work that addresses solely specific content, such as database-related methods or web page content, our approach can be applied to cache results of any computation done by a web application, which includes complex logic and personalized web content. Alternatively, it can be used as a decision-support tool to help developers in the process of deciding what to cache, guiding them while manually implementing caching.

In summary, we provide the following contributions: (i) a caching approach focused on integrating caching into web applications in a seamless and straightforward way, providing an automated and adaptive management of cacheable methods; (ii) a framework that detaches caching concerns from application code; and (iii) an empirical evaluation of our automated approach that indicates that it can effectively identify cacheable opportunities in web-based applications.

We next provide background on application-level caching. We detail the proposed approach in Section~\ref{sec:approach}, and give details of its implementation as a framework in Section~\ref{sec:framework}. An empirical evaluation of our approach is presented in Section~\ref{sec:evaluation}. Limitations and related work are discussed in Sections~\ref{sec:limitations} and~\ref{sec:related_work}, respectively. Finally, we conclude in Section~\ref{sec:conclusion}.

\section{Background on Application-level Caching}
\label{sec:background}


\subsection{Overview}

\begin{figure}
    \centering
    \includegraphics[width=0.7\textwidth]{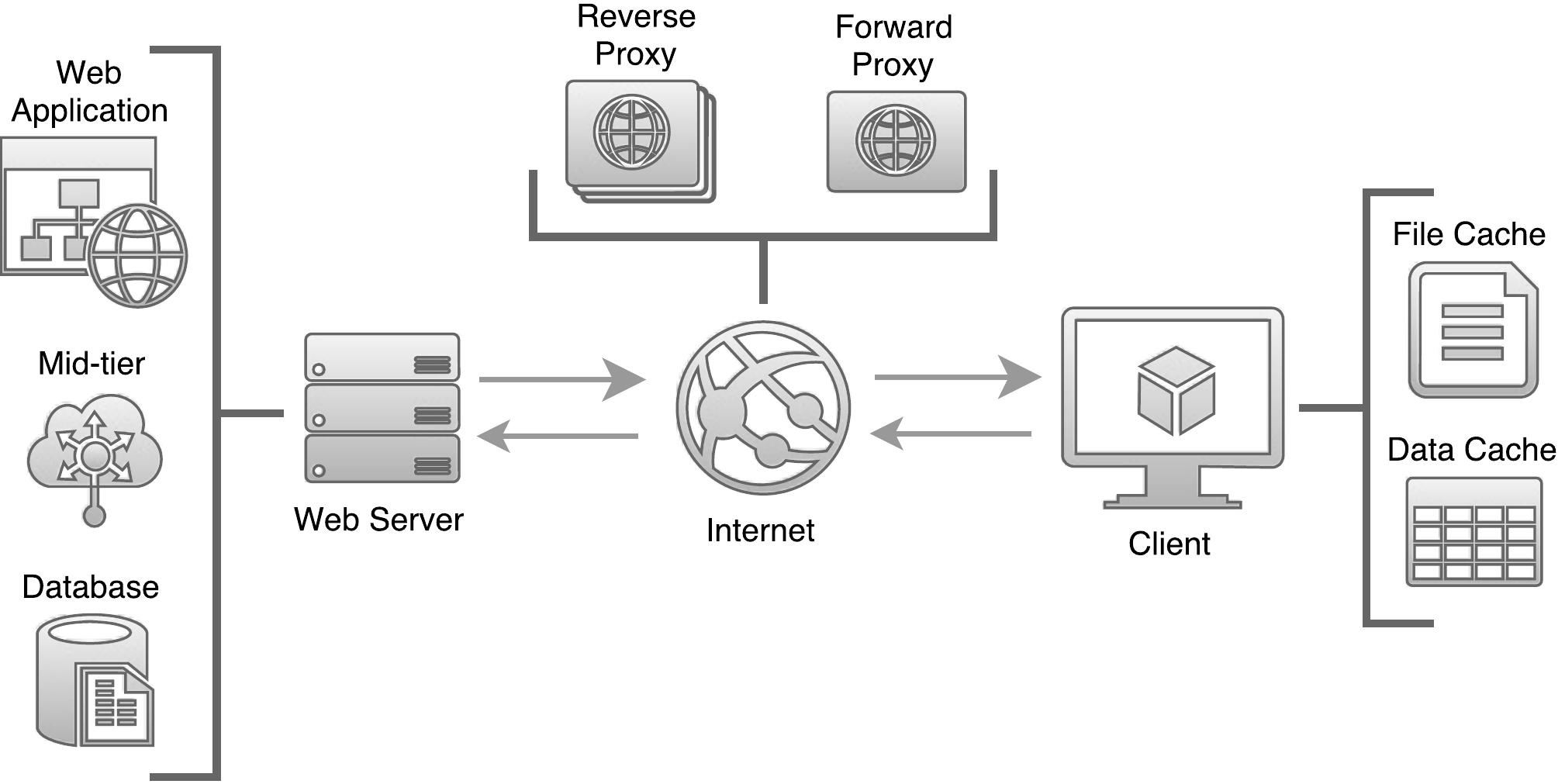}
    \caption{Traditional Web Application Architecture with Associated Caching Locations.}
    \label{fig:webinfrastructure}
\end{figure}

Given that web caching is strictly related to the web infrastructure components, we overview such architecture in Figure~\ref{fig:webinfrastructure} together with the primary caching locations. Caching solutions at the web infrastructure can be grouped into three main locations: client, Internet, and server~\cite{Labrinidis2009,Ravi2009,Mertz2017a}. In short, the \emph{client} component is essentially the client's computer and web browser; while the \emph{Internet} component contains a wide range of different, interconnected mechanisms to enable the communication between client and server. The \emph{server} can include multiple and different servers that are collectively seen as the web server by the client~\cite{Mertz2017a}.

Each caching location has its benefits, challenges, and issues that together lead to trade-offs to be resolved when choosing a caching solution. For instance, particular forms of content can be cached according to a selected choice of location as well as the abstraction level provided, which can be fully transparent to the application or tightly integrated into the application code. In addition, hit and miss probabilities vary across different locations~\cite{Negrao2015,Li2006a,Guerrero2011,Ports2010,Amza2005,Baeza-Yates2007}. Due to the variety of application domains, workload characteristics, and access patterns, no universal web caching solution outperforms other caching options in all possible scenarios. Therefore, caching at these different locations is complimentary. 

Application-level caching resides on the server side. As opposed to traditional caching alternatives, which are transparent to the application, application-level caching is tightly integrated into the application base code. It becomes needed in modern web applications because they manipulate and process customized content and, in this case, caching final web pages provides limited benefits. Therefore, application-level caching can be used to separate generic from customized content at a fine-grained level.

\begin{figure}
  \centering
  \subfigure[Code Example.]{\includegraphics[width=0.45\linewidth]{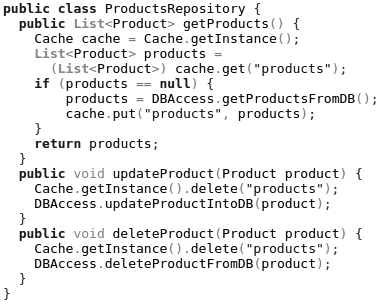}
  \label{fig:codeExample}}
  \subfigure[Steps of the Caching Process.]{\includegraphics[width=0.45\linewidth]{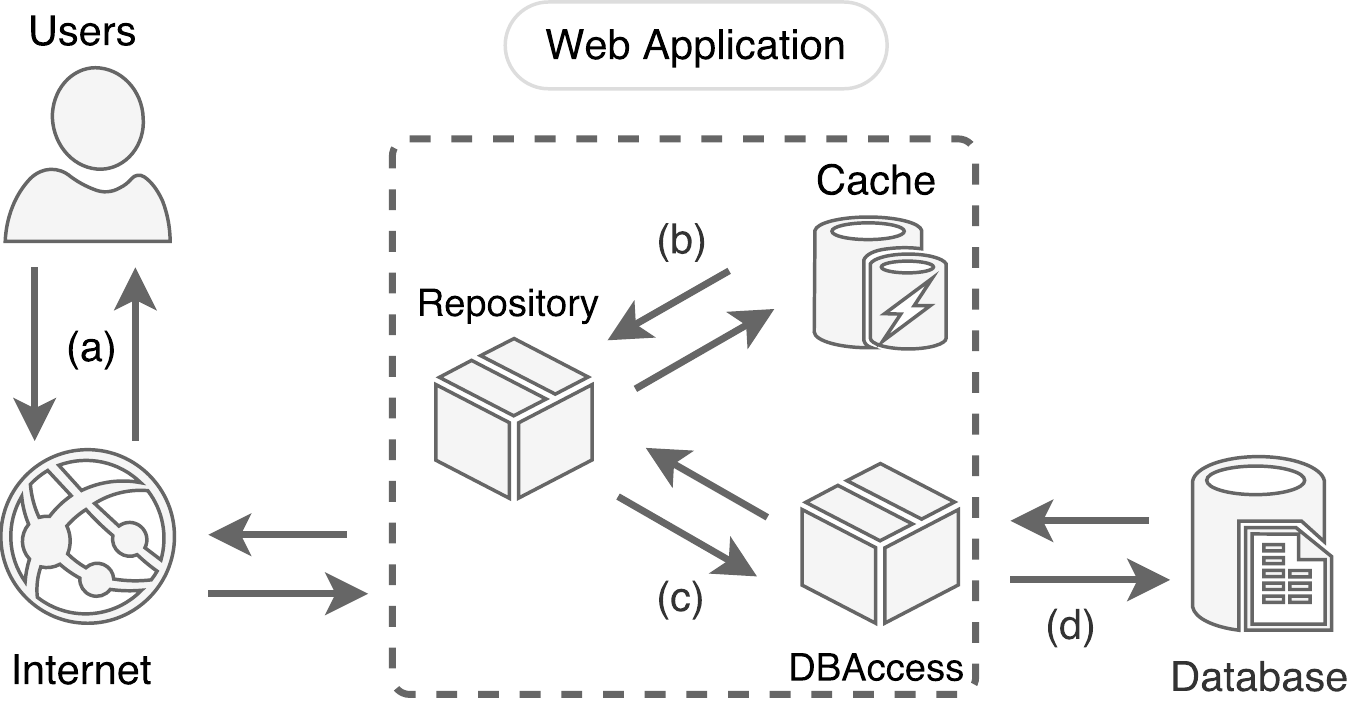}
  \label{fig:applicationcaching}}
  \caption{Application-level Caching Overview.}
  \label{fig:AppCacheOverview}
\end{figure}

As an example, we present in Figure~\ref{fig:AppCacheOverview} a scenario in which application-level caching is used to lower the database workload. In this example, an e-commerce application has a \texttt{ProductsRepository} class, which is responsible for loading products from the database. First, the web application receives a request for all the products from a user (\emph{step a}), which eventually leads to an invocation to the \texttt{ProductsRepository} class. However, such class makes a database query on \texttt{DBAccess}, and calling and executing \texttt{DBAccess} may imply an overhead regarding computation or bandwidth. Therefore, \texttt{ProductsRepository} manages to cache \texttt{DBAccess} results and, for every request, \texttt{ProductsRepository} verifies whether \texttt{DBAccess} should be called or there are previously computed results already in the cache that can be used (\emph{step b}). Then, the cache component performs a look up for the requested data and returns either the cached result or a not found error. If a \emph{hit} occurs, it means the content is cached, and \texttt{ProductsRepository} can avoid calling \texttt{DBAccess}. However, when a \emph{miss} occurs and then a not found error is returned, it means that \texttt{DBAccess} computation is required (\emph{steps c} and \emph{d}). The newly fetched result of \texttt{DBAccess} can then be cached to serve future requests faster. These steps, described in Figure~\ref{fig:applicationcaching}, are those typically performed in any cache implementation. The key difference is that, in application-level caching, the responsibility of managing the caching logic is entirely left to application developers, who must manually handle the cache.

In this scenario, caching decisions are made explicit, which involves many issues~\cite{Mertz2016}. The first is that developers must manually develop and insert the caching logic into the application base code, which involves content retrieval and translation as well as key assignment and consistency maintenance. Such logic is usually tangled with the business logic---as illustrated in Figure~\ref{fig:codeExample}---and making the caching code a \emph{cross-cutting concern}, i.e.\ it is spread all over the application base code, resulting in increased complexity and maintenance time~\cite{Ports2010}.

Furthermore, such implementation requires developers to make caching decisions, such as choosing which objects to get, put or remove from the cache. Such decisions demand a significant effort and reasoning from developers because they need to understand what are the typical usage scenarios, how often the content selected to be cached is going to be requested, how much memory it consumes, and how often it is going to be updated. These are all \emph{non-trivial decisions}~\cite{Mertz2016}.

Caching implementation can be supported by available libraries and frameworks, which provide implementations of a cache system, e.g.\ Caffeine\footnote{\url{https://github.com/ben-manes/caffeine}}, EhCache\footnote{\url{http://www.ehcache.org/}} and Memcached\footnote{\url{https://memcached.org/}}. Although there are existing tool-supported approaches that raise the abstraction level of caching and prevent adding much cache-related code to the base code~\cite{Gupta2011,Ports2010}, caching decisions such as determining \emph{what should be cached} remains as a developer responsibility.

A fundamental problem of application-level caching is that all issues above usually demand extensive knowledge of the application to be properly solved. Consequently, developers manually design and implement solutions for all these mentioned tasks. However, even when adopting a caching solution to improve the application performance, the issues and challenges concerning maintenance remain unaddressed. While designing and implementing application-level caching, developers usually specify and tune cache configurations and strategies according to application specificities. Nevertheless, an unpredicted or unobserved usage scenario may eventually emerge. As the cache is not optimized for such situations, it would likely perform sub-optimally~\cite{Radhakrishnan2004}. As a result, to achieve caching benefits so that the application performance is improved, it is necessary to tune cache decisions continuously~\cite{Santhanakrishnan2006}.

This shortcoming motivates the need for \emph{adaptive caching solutions}, which could overcome these problems by automatically adjusting caching decisions according to the application specificities to maintain a required performance level. Moreover, an adaptive caching solution minimizes the challenges faced by developers, requiring less effort and providing a better experience with caching.

\subsection{Selection of Cacheable Content}

Although application-level caching is commonly being adopted, the selection of cacheable content is typically an ad hoc and empirical process. To find caching best practices, developers can make use of widespread knowledge, consult development blogs, or simply search online for tips. Nevertheless, this knowledge is unsupported by concrete data or theoretical foundations that demonstrate its effectiveness in practice. Thus, developers usually implement the necessary cache logic for the assumed cache opportunities and evaluate the performance improvement empirically based on benchmarks and application executions.

\begin{figure}
\centering
\includegraphics[width=0.7\linewidth]{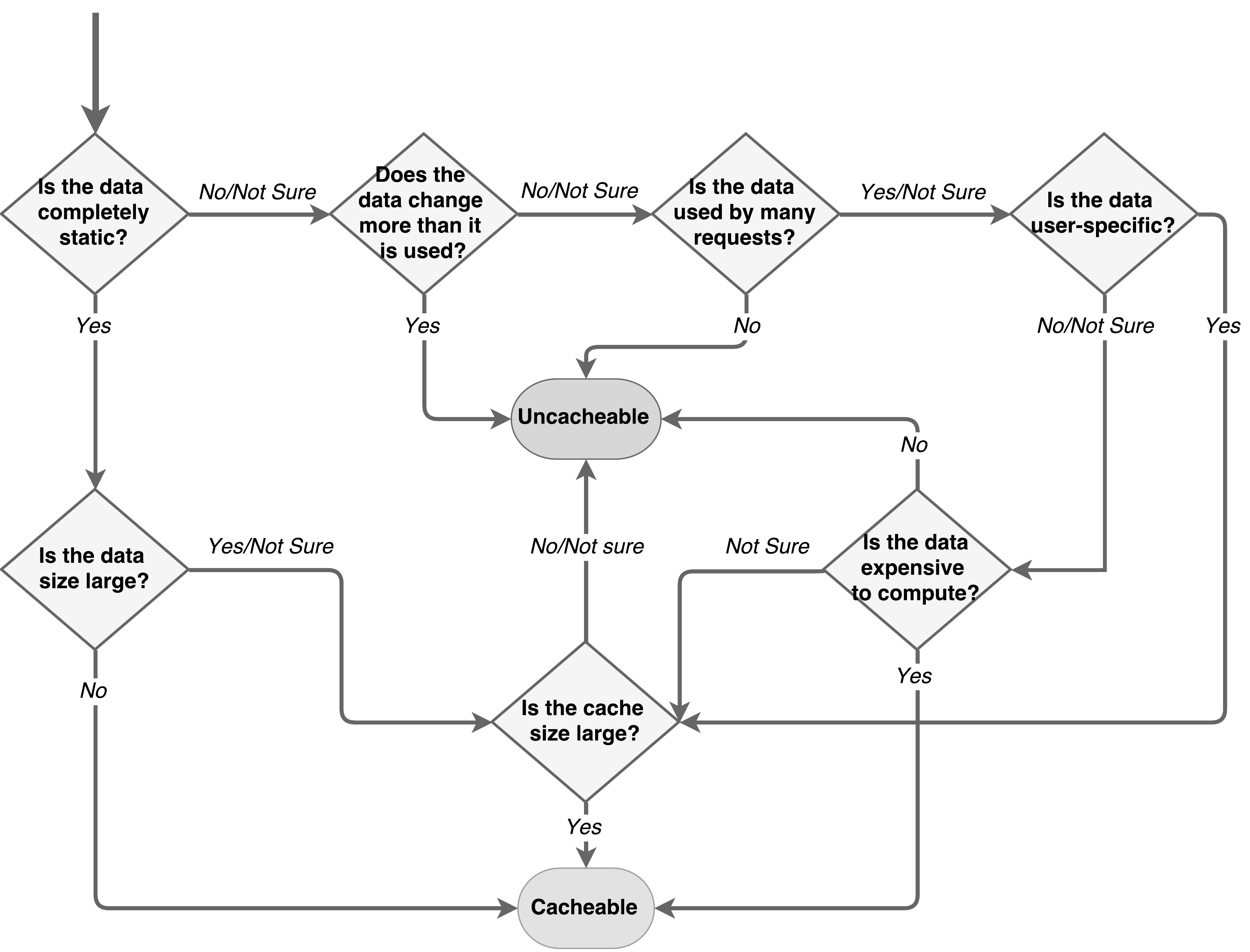}
\caption{Cacheability Pattern~\cite{Mertz2016}.}
\label{fig:cachebility}
\end{figure}

Given this scenario, in previous work~\cite{Mertz2016}, we analyzed how developers deal with application-level caching. This work allowed us to derive a set of patterns that capture criteria used to make cache-related decisions, giving practical guidelines for developers to appropriately design caching in their applications. One of such patterns, namely the \emph{Cacheability Pattern}, focuses on the selection of cacheable content, more specifically \emph{method calls}, and is specified in a flowchart, presented in Figure~\ref{fig:cachebility}.

Although such pattern is conceived to reuse, it is an \emph{abstract} solution that requires specific reasoning and coding to make use of it. In real-world situations, developers should reason about application specificities and decide whether to cache or not a specific method regarding the criteria. \emph{No objective measurement to evaluate whether a method satisfies each criterion has been provided.} Therefore, although such an approach systematizes the decision regarding content cacheability, developers must still have knowledge of the application requirements, workload, domain, access patterns and business logic as well as manually select content to be cached.

\section{Automated Application-level Caching Approach}
\label{sec:approach}


Based on the structured and documented knowledge captured by the Cacheability Pattern, we propose an approach that can automatically make decisions regarding cacheable content. In this section, we first present an overview of our approach. Next, we detail each of its key activities.

\subsection{Approach Overview}

Our approach is based on the decision process presented in Figure~\ref{fig:cachebility}, which gives the criteria to be taken into consideration to make caching decisions. For each criterion, we propose \emph{objective means} of measuring application methods and evaluating whether they should be cached. This evaluation is based on data collected by monitoring the workload of web applications at runtime. Furthermore, our automated approach takes into account application-specific details in caching decisions, which is, in fact, the information considered by developers in the development of a caching solution, thus providing a solution that is specific and optimized to the problem of selecting cacheable content. By providing an automated approach, we relieve developers from the process of instantiating that abstract solution according to their specific domains or characteristics.

\begin{figure}
\centering
\includegraphics[width=\linewidth]{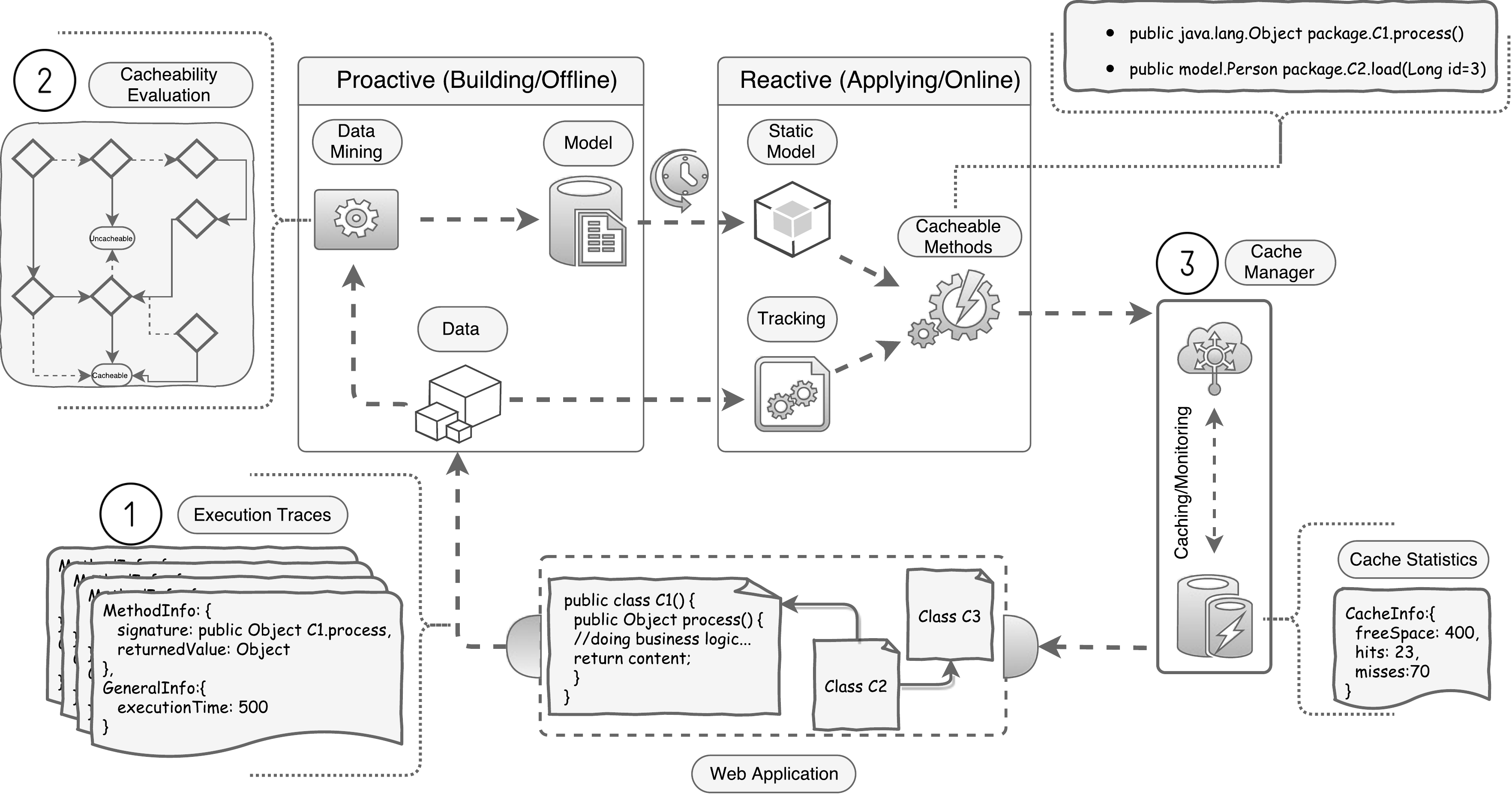}
\caption{Overview of our Application-level Caching Framework.}
\label{fig:proposedsolution}
\end{figure}

Our proposed solution can be incorporated into web applications so that it can monitor and choose content to be cached according to changing usage patterns. It consists of two complementary asynchronous parts: (a) a \emph{reactive model applying}, responsible for monitoring traces of the application execution and caching method calls (previously identified as caching opportunities) at runtime; and (b) a \emph{proactive model building}, which analyzes on the background the behavior of the application, taking into account application-specific information, and finds cacheable opportunities. Figure~\ref{fig:proposedsolution} presents an overview of the dynamics of our approach, indicating the activities that comprise its running cycle. These asynchronous parts involve performing three different activities: (1) data tracking, (2) data mining and (3) cache management.

Because our approach monitors the application at runtime and self-optimize caching decisions, it is sensitive to the evolution of application workload and access patterns. This addresses the main problem regarding cache maintenance because it forces developers to revise their cache decisions constantly. Moreover, the integration between our approach and the application is seamless and does not require manual inputs, detaching caching concerns from the application and providing higher cohesion and lower coupling. As result, our approach can reduce the reasoning, time and effort required from developers to develop a cache solution, allowing them to dedicate time to write the most relevant code (i.e.\ business logic). If full automation is not desired, our approach can alternatively serve as a decision-support tool to help developers in the process of deciding what to cache, guiding them while manually implementing caching.

We next conceptually describe each activity of our approach. Later, we show how they are operationalized within an implemented framework.

\subsection{Data Tracking: Monitoring Execution Traces}

As opposed to traditional caching approaches, which cache content such as web pages or database queries, our approach focuses on caching \emph{method-level content}. Moreover, we use application-specific information to make caching decisions, such as the user session, execution time of method calls, and data and cache sizes.

To monitor the application behavior, we collect \emph{method invocations or calls} (i.e.\ application execution traces) at runtime. Related to this monitoring process, two issues must be addressed. First, we must specify what information should be recorded when invoking methods. Second, given that the monitored and recorded information may be a complex structure, it is essential to provide means of dealing with such complexity.

Concerning the first issue, we adopt a \emph{lightweight} and \emph{conservative} approach. It is lightweight because it is based on recording just the input and output of each method call. As stated by Toffola et al.~\cite{DellaToffola2015}, recording detailed information before and after each method call does not scale to large web applications with a high number of concurrent users. Thus, we focus on information regarding the arguments passed to the call and the return value of the call (if any), because this information is sufficient to describe the relevant input and output state for most of the methods. The monitoring process is also conservative in the sense that the recorded information is the complete method call, consisting of the method identification (i.e.\ its signature), the values of all method inputs, its output (i.e.\ its returned value), and additional information, namely cost and user session. 

Each recorded method call is a tuple \[\langle s, P, r, c, u\rangle\] where $s$ is the representation of the target of the call, $P = [ p_1, \dots, p_n ]$ is a list of parameters of the call, $r$ is the returned value, $c$ is the cost of computing the method, and $u$ is the user session associated with the method call. The cost can be the time taken to execute the method, memory consumed, or any other developer-specified resource.

\begin{figure}
\centering
\includegraphics[width=\linewidth]{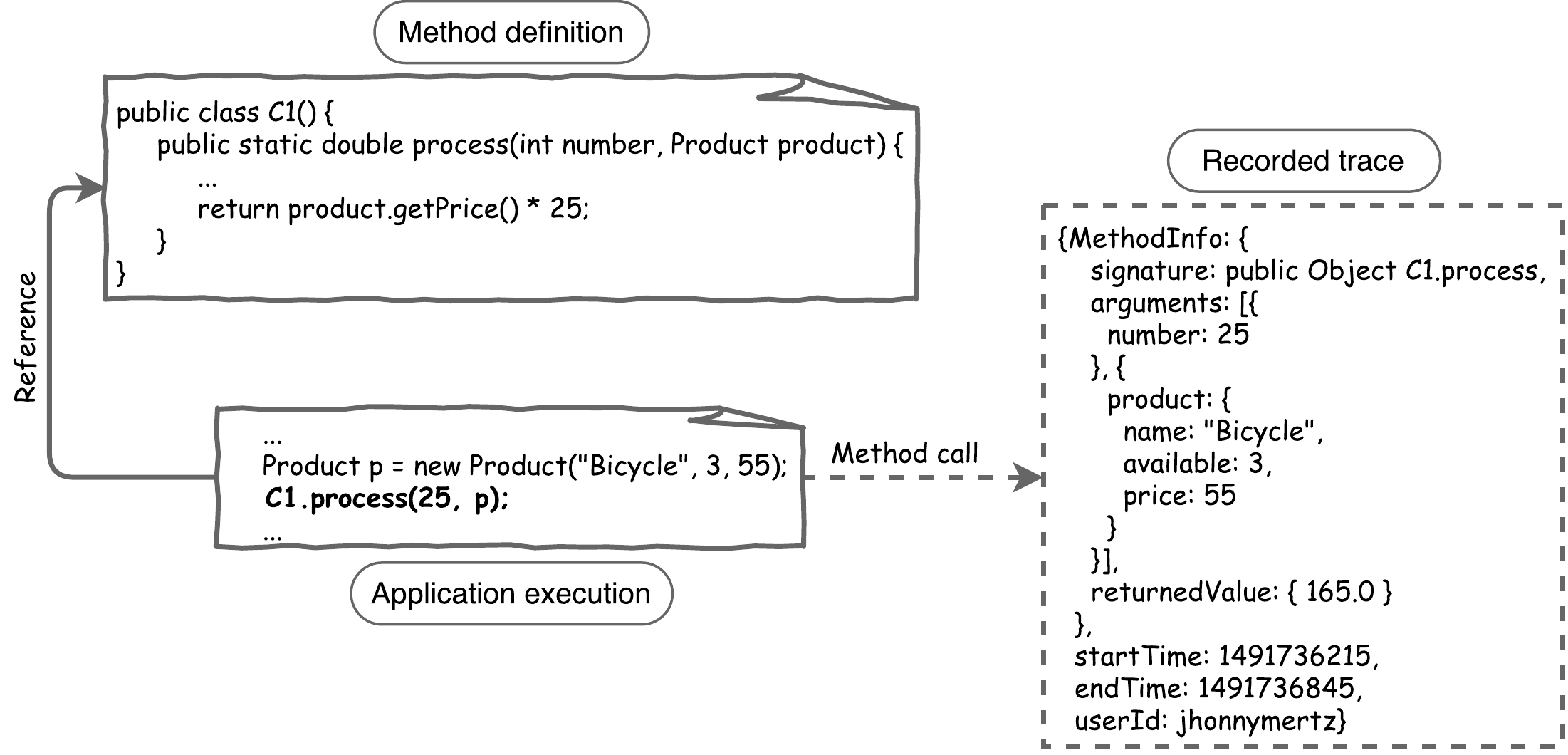}
\caption{Example of a Method Calls translated to an Execution Trace.}
\label{fig:trace-example}
\end{figure}

Finally, these method calls are mapped into a generic representation that is saved as a string. The representation describes the data itself and the shape of the data item. The representation captures the structure of objects and is independent of the memory locations where objects are stored or other globally unique identifiers of objects. Figure~\ref{fig:trace-example} gives an example of how a method call is mapped into a record.

\subsection{Data Mining: Identifying Cacheable Method Calls}

The second activity of our approach takes as input the output of the previous activity (data tracking), which provides us with information needed to identify methods to be cached. To decide what should be cached, the second activity is based on the mining of such information.

As said, the reasoning part of our automated approach is based on the decision process presented in Figure~\ref{fig:cachebility}. This reasoning model specifies a sequence of questions to be answered that are associated with criteria to be analyzed. By chaining different decisions taking into account each criterion, an importance relationship among them is established. Content changeability is the first analyzed criterion, followed by usage frequency, shareability, retrieval complexity, and size properties. However, given the pattern was conceived to be used at design time, using it at runtime requires adaptations to be made. 

These adaptations concern the size-related criteria, which make more sense at design time, because developers may not have enough information to predict the size required to store a particular piece of content and the size available in the cache. Consequently, in this case, some data may be chosen as not cacheable. However, at runtime, it is possible to know the size of the content to be cached as well as the available cache size and occupation ratio.

\begin{figure}
\centering
\includegraphics[width=0.7\linewidth]{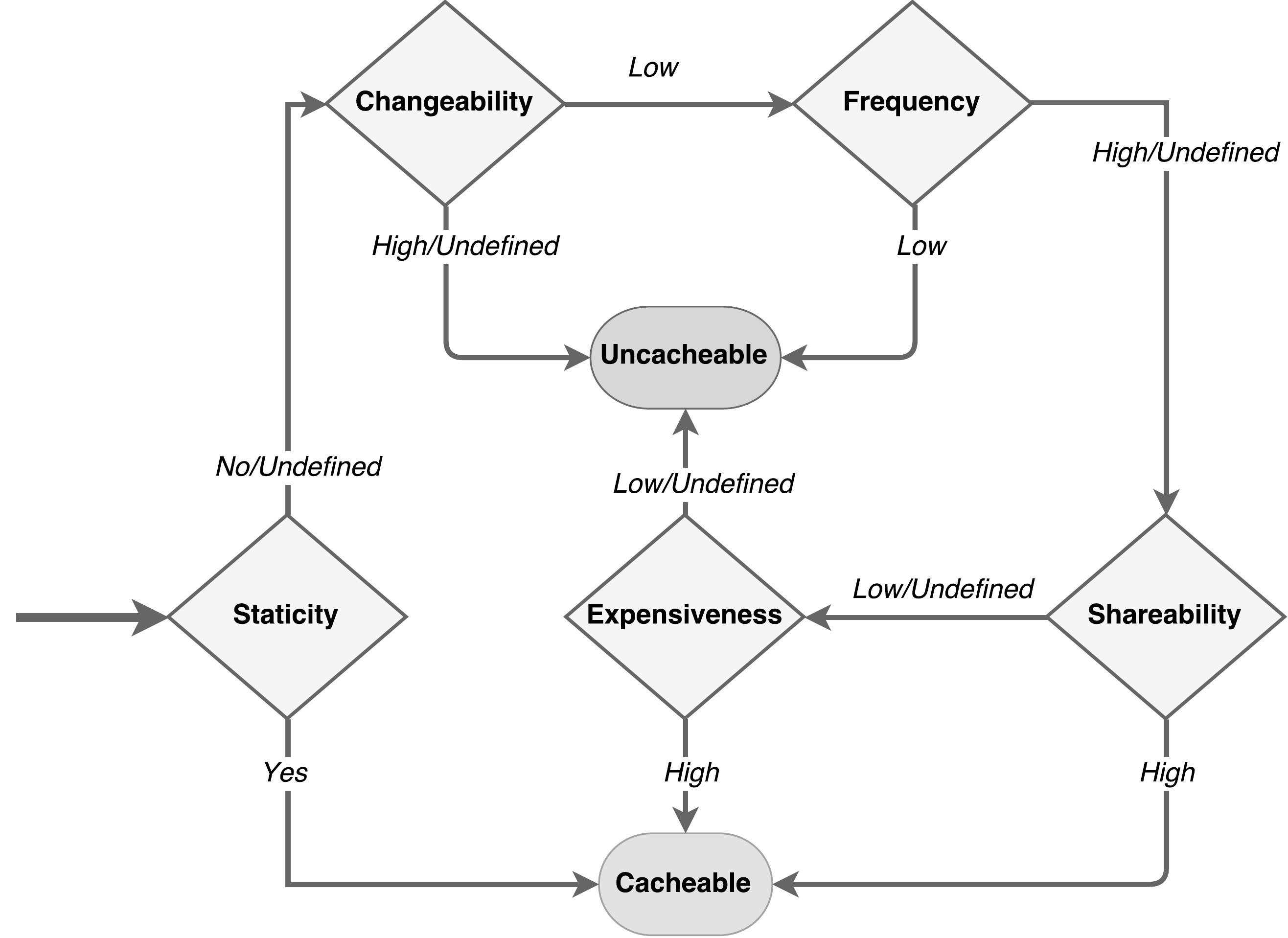}
\caption{Criteria associated with the Cacheability Pattern.}
\label{fig:cachebility_metrics}
\end{figure}

Therefore, using size-related criteria to decide, at runtime, what to cache (as suggested by the Cacheability pattern) can lead the cache to a non-maximum utilization, i.e.\ cacheable methods could be classified as uncacheable even with enough free space in the cache. To avoid this, we use the size-related information to decide whether a cacheable method should be put in the cache. Consequently, we do not consider the size-related criteria in the identification of cacheable method calls in the data mining activity, but they are taken into account in the next activity. As result, we obtain a simplified reasoning process to be automated and used while identifying cacheable content, which is presented in Figure~\ref{fig:cachebility_metrics}.

To be automated, this process needs to have its decision criteria analyzed to make decisions. However, in the Cacheability Pattern, there are no \emph{objective} definitions. This means that, when adopting this pattern, developers must provide a meaning for each criterion. We thus, as part of our approach, propose objective measurements to evaluate each criterion. There are five specified measurements, which are detailed in Table~\ref{tab:criteria}.

\begin{table}
    \centering
    \scriptsize
    \caption{Objective Evaluation of the Cacheability Pattern Criteria.}
    \label{tab:criteria}
    \begin{tabular}{p{2.2cm} l p{9.2cm}}
    \toprule
    \textbf{Question} & \textbf{Criterion} & \textbf{Meaning} \\ \midrule 
      Is the data completely static? & Staticity & A method staticity is associated with how many times a method returns the same value when it receives the same parameters. Staticity is given by $staticity(m) = |P_{Set}| / |Pr_{Set}|$, where $P_{Set}$ is the set of different lists $P$ of parameter values received by a method $m$, and $Pr_{Set}$ is the set of different tuples $\langle P, r \rangle$, where $P$ is a list of parameters values and $r$ is the returned value. A method $m$ is said to be \textbf{completely static} if $staticity(m) = 1$.\\ \hline
      
      Does the data change more than it is used? & Changeability & Static methods tend to achieve the highest hit ratio when cached. However, caching methods that often do not change can still bring benefits. Therefore, the changeability of a method is the dual of staticity, i.e.\ $changeability(m) = 1 - staticity(m)$. To evaluate whether a method does not often change, we use as a reference value $\mu_{ch} + k \times \sigma_{ch}$, where $\mu_{ch}$ and $\sigma_{ch}$ are the changeability mean and standard deviation, respectively, and $k$ is a given number. The changeability criterion has a yes answer when the method changeability is below the reference value, i.e.\ when it is $k$ standard deviations below the changeability mean. \\ \hline
      
      Is the data used by many requests? & Frequency & A method frequency is associated with how many times a method is called, and this criterion is used in our approach also to assess whether the collected trace sample is large enough to make decisions. Therefore, we use a specified threshold to distinguish frequent from unfrequent methods. The threshold is the sample size $size(c, e)$, where $c$ is a confidence level e $e$ is a margin of error. If the number of collected traces of a method is above the required sample size, it is said to be \textbf{frequent}. \\ \hline
      
      Is the data user-specific? & Shareability & The method shareability gives how much the results of a method call are shared among different users because if results of a method are shared among many users, caching this method may potentially increase the hit ratio. A method shareability $shareability(m)$ is the percentage of different user sessions in which requests lead to a method call with the same parameter values. Similarly to frequency, a method is said \textbf{shared} if its shareability is $k$ standard deviations $\sigma_{sh}$ above the shareability mean $\mu_{sh}$, that is, $shareability(m) > \mu_{sh} + k \times \sigma_{sh}$. Anonymous method calls (not associated with any user) are not taken into consideration in this criterion. \\ \hline
      
      Is the data expensive to compute? & Expensiveness & A method expensiveness is associated with the cost $cost(m)$ for computing it, which can be the time taken to compute it or consumed memory, for example. As above, a method is said \textbf{expensive} if $cost(m) > \mu_{ct} + k \times \sigma_{ct}$, where $\mu_{ct}$ and $\sigma_{ct}$ are the cost mean and standard deviation, respectively. \\
    \bottomrule
\end{tabular}
\end{table}

Essentially, the objective evaluation is a statistical analysis of collected traces. All the information required by each criterion is presented in Table~\ref{tab:infos}. Note that due to a limited amount of data---such as considering a method with only a few executions as static, frequent, or less changing---wrong conclusions can be reached. Therefore, in situations in which we have an insufficient amount of data, we assume \emph{Undefined} as the result of the criterion analysis.

\begin{table}
    \centering
    \scriptsize
    \caption{Required Information for Evaluating Cacheability.}
    \label{tab:infos}
    \begin{tabular}{ll}
    \toprule
    \textbf{Criterion} & \textbf{Required information} \\ 
      Staticity & Input and output of the method calls \\
      Changeability & Input and output of the method calls \\
      Frequency & Input and output of the method calls \\
      Shareability & Session user that lead to the method calls \\
      Expensiveness & Execution time of the method calls \\
    \bottomrule
\end{tabular}
\end{table}

Based on our objective criteria evaluation and the simplified decision process, our approach identifies cacheable methods. It is important to note that one method may result in many cacheable opportunities (and consequently many entries in the cache) because our approach distinguishes and analyzes method calls, which are specified as a combination of the method signature and parameter values. In the next activity, described as follows, we detail how we cache method calls defined as cacheable opportunities, using the two remaining criteria, not taken into account in this activity.

\subsection{Cache Management: Caching Identified Opportunities}

With the previous activities, we identify a set of cacheable method calls. We now discuss how our approach manages the cache component to cache the selected method calls as well as to keep consistency. Similarly to the data tracking activity, through monitoring the application execution, we intercept calls to cacheable methods and assess whether the content associated with the call is in the cache.

As our approach solely learns whether method executions should be cached, other cache concerns, such as eviction and consistency, were addressed with standard solutions. To make our approach less dependent on the effectiveness of alternative cache policies and algorithms, as a default configuration, we suggest a conservative approach that caches content only when there is enough space in the cache, considering the data size and cache size, the two remaining decisions of the Cacheability pattern, which are presented in Table~\ref{tab:size_criteria}. Moreover, to periodically free space in the cache and remove outdated content, a time-to-live (TTL) should be specified. With TTL, cached methods expire after a time in cache, regardless of possible changes. This frees cache space for caching new content associated with cacheable methods. In addition, TTL is a popular solution to deal with consistency issues, which requires to relax freshness and admit potential staleness to increase performance and scalability.

\begin{table}
    \centering
    \scriptsize
    \caption{Objective Evaluation of the Size-related Criteria.}
    \label{tab:size_criteria}
    \begin{tabular}{p{3cm} p{2cm} p{7cm}}
    \toprule
    \textbf{Question} & \textbf{Criterion} & \textbf{Meaning} \\ \midrule
      Is the cache size large? \par Is the data size large? & Cache size \par Data size & A method is only considered to be cached if $size(m) < free(c)$, where $size(m)$ and $free(c)$ are the estimated size of the result of a method call and the available space in the cache, respectively. \\
    \bottomrule
\end{tabular}
\end{table}

During the execution, when a cacheable method is called and the returned content is not in the cache, we estimate how much space of the cache this method call requires and verify whether the cache has the corresponding free space to allocate such content. If there is no enough space in the cache, no content is cached until TTL expires cached data. It is important to note that cache size and TTL are both domain-specific and have no relation to the identification of cacheable content. Thus, these values should be manually specified when instantiating our approach.

\section{\fw{} Framework}\label{sec:framework}

\begin{figure}
\centering
\includegraphics[width=0.7\linewidth]{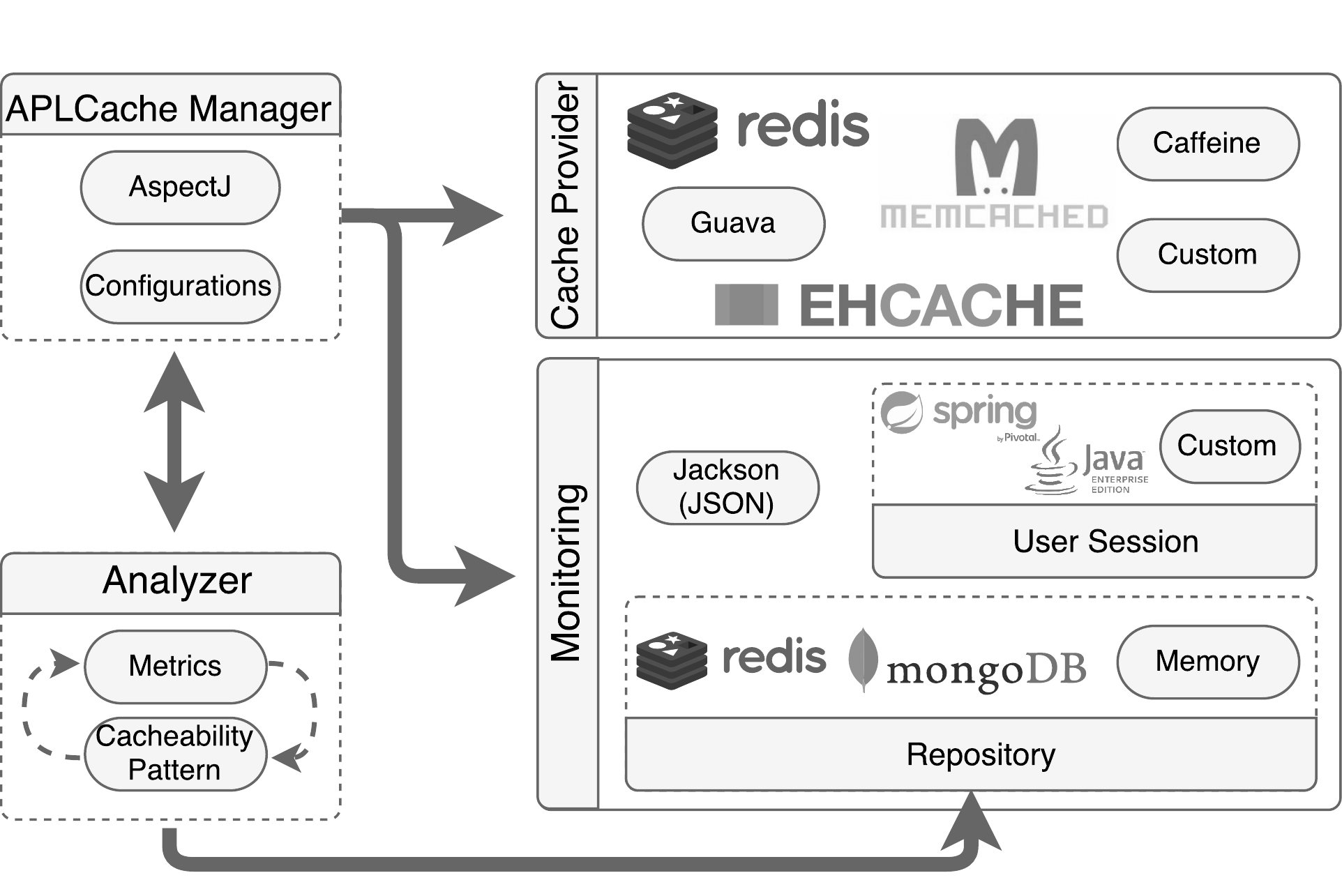}
\caption{\fw{} Architecture and Technologies.}
\label{fig:framework}
\end{figure}

The three main activities of our approach, conceptually described above, were implemented as a framework, namely \fw{}, that can be instantiated to integrate web applications. Our framework is implemented in Java, thus can be used with Java web applications. This choice is due to our previous programming experience and tools available that were adopted as part of our implementation. Moreover, we adopted a set of technologies that provide an appropriate infrastructure for the framework. Used technologies are highlighted in Figure~\ref{fig:framework}, which presents the \fw{} architecture, with its modules and communication among them.

To collect data to be analyzed and manage cacheable methods, we intercept method executions using aspect-oriented programming (AOP), more specifically the AspectJ\footnote{\url{https://eclipse.org/aspectj/}} implementation. AOP provides an easy way to weave code at specified points, without changing the base code. Considering the explained generic representation to save application traces, to compare input and output data of method calls and ensure that structurally equivalent objects have the same representation, objects are loaded and compared by using implementations of the \emph{equals} and \emph{hashCode} methods. Saving actions are performed in an execution thread separate from the one that is processing the request, minimizing response delays. Our framework also provides means for developers to make customizations using hints, indicating possible locations of cacheable methods, which can improve the set of caching candidates as well as exclude methods that should not be cached, thus saving the time of tracking them.

Available solutions of caching components were also adopted. \fw{} automates the decision of what to cache, while these caching components provide APIs that allow us to manipulate data and access metadata information about the cache, such as statistics and configurations. Our framework is decoupled from particular caching components, and thus supports the most popular distributed cache systems and libraries, which can be configured through property files and annotations. Therefore, \fw{} provides a fully customizable environment where different cache policies and algorithms can be configured and used along with the proposed approach to detect cacheable content. Nevertheless, although we provide an off-the-shelf solution, we strongly encourage developers to customize and tune components according to their needs and preferences to achieve improved results.

The collected data are analyzed offline, separately from the web application, to prevent impact in the application performance---it can even run on a dedicated machine. To evaluate shared execution traces (accessed by multiple users), we obtain the user session, which is an application-specific information thus taken into account only in application-level caching. Our framework provides a set of alternative implementations to obtain this information from the most popular web frameworks, such as Java EE and the Spring Framework\footnote{\url{https://spring.io/}}. In case alternative ways of managing user sessions are adopted, developers should implement interfaces provided by our framework.

\fw{} not only can be used to incorporate our proposal to web applications but was also used in our evaluation, which is presented next.



\section{Evaluation}
\label{sec:evaluation}


In this section, we proceed to the evaluation of our automated caching approach, measuring two aspects: (i) performance; and (ii) differences between developers' caching decisions and our approach. Given that collecting data to make caching decisions is required either if developers manually analyze such information, or automatically by an algorithm, our evaluation focuses on the caching decision rather than the monitoring process. We first describe our evaluation procedure and then discuss obtained results and threats to validity.

\subsection{Goal and Research Questions}
\label{sec:goalandquestions}

As stated in the introduction, our primary objective while proposing this approach is to provide automation and guidance to developers when adopting application-level caching in their applications. This approach aims to provide such guidance using an \emph{automated and seamless application-centric approach}, i.e.\ by identifying cacheable opportunities in their applications, taking into account application details. The evaluation of our approach is based on the goal-question-metric (GQM) approach proposed by Basili et al.~\cite{Basili1994}. The GQM was adopted to define the goal of the evaluation, the research questions to be answered to achieve the goal and metrics for responding to these questions. Following this approach, we present in Table~\ref{tab:gqm} the description of the evaluation, following the GQM template. To achieve our goal, we investigated different aspects of our automated approach, which are associated with three key research questions presented in Table~\ref{tab:rqs} along with their metrics.

\begin{table}
    \centering
    \scriptsize
    \caption{Goal Definition (GQM Template).}
    \label{tab:gqm}
    \begin{tabular}{p{4cm} p{8cm}}
    \toprule
    \textbf{Definition Element} & \textbf{Evaluation Goal} \\ 
    \midrule
    Motivation                  & To assess the improvements provided by our automated application-level caching approach, \\
    Purpose                     & evaluate \\
    Object                      & the effectiveness of the automated caching approach \\
    Perspective                 & from a perspective of the researcher \\
    Domain: web-based applications     & when compared to manually developed caching solutions \\ 
    Scope                       & in the context of 3 software projects, obtained from open-source repositories. \\
    \bottomrule
\end{tabular}
\end{table}

\begin{table}
    \centering
    \scriptsize
    \caption{Research Questions and Metrics.}
    \label{tab:rqs}
    \begin{tabular}{p{5.6cm} p{7cm}}
    \toprule
    \textbf{Research Question} & \textbf{Metric} \\ 
    \midrule
    \textbf{RQ1.} What is the performance improvement provided by caching method calls selected by our automated caching approach? & 
      \textbf{M1-1.} Throughput.
      \par \textbf{M1-2.} Hit ratio.
      \par \textbf{M1-2.} Total number of cache hits.
    \\ \midrule
    \textbf{RQ2.} What are the similarities and differences between decisions made by our automated approach and by developers? & 
      \textbf{M2-1.} Number of caching opportunities that were identified by both developers and our automated approach.
      \par \textbf{M2-2.} Number of caching opportunities identified only by developers.
      \par \textbf{M2-3.} Number of caching opportunities identified only by our automated approach.
      \\ 
    \bottomrule
\end{tabular}
\end{table}

Given that caching is essentially a technique to improve performance and scalability, RQ1 concerns evaluating this. Although our approach relieves the developer from this task, it still needs to provide performance improvements. Determining the cacheable content and the right moment of caching or clearing the cache content are a developer's responsibility and might not be trivial in complex applications. In RQ2, we aim to identify what methods were considered caching opportunities and how they can be compared to the choices manually made by developers.

\subsection{Procedure}

To evaluate our approach, we used performance test suites, which aim to simulate real-world workloads and access patterns~\cite{Binder2000}. Furthermore, these tests have been used to evaluate improvements of caching in web applications~\cite{DellaToffola2015,Chen2016}. 

Simulations were performed using three different caching configurations: (i) no application-level caching (\textbf{NO}), (ii) application-level caching manually designed and implemented by developers (\textbf{DEV}); and (iii) our approach (\textbf{AP}). To assess performance, we used three metrics: \emph{throughput} (number of requests handled per second), \emph{hit ratio} and the \emph{total number of hits}, because they are well-known in the context of web applications and cache performance tests.

Our simulation emulated client sessions to exercise applications and evaluate decision criteria. Simulations consisted of variations of 1, 5, 10, 25 and 50 simultaneous users constantly navigating through the application, always at a limit of 500 requests per user. Each simulation was repeated ten times, and the mean of each metric was collected. Each emulated client navigates from an application page to another, selecting the next page from those accessible from the current page, to better represent a real user. The navigation process starts on the application home page and follows a non-uniform random selection that falls into a distribution where 80\% of the requests are read-only, while the remaining 20\% perform at least one write operation. This distribution is mentioned in standard performance benchmarks such as TPC-W\footnote{\url{http://www.tpc.org/tpcw/}} and RUBiS\footnote{\url{http://rubis.ow2.org/index.html}}.

The evaluation of different criteria requires different parameters. We used a web application, not used in our evaluation, to empirically choose these parameters. As result, we adopted 99\% and 3\% as the confidence level and margin of error, respectively, for the frequency criterion. For shareability and expensiveness, we adopted $k = 1$, while for changeability, $k = 0$. For expensiveness, the cost corresponds to the method execution time.

For all the executions of AP, caches were bounded by a cache size and configured with a TTL, which were both specified and used by developers of our target systems (DEV). As the results may be influenced by the cache policy adopted, in addition to our standard technique based on TTL, we also evaluated our approach combined with popular replacement policies, namely least recently used (LRU) or least frequently used (LFU). Finally, we used information collected over 2 minutes to build the caching decision model in AP scenario.

Our evaluation was performed with three open-source web applications\footnote{Available at \url{http://www.cloudscale-project.eu/}, \url{https://github.com/SpringSource/spring-petclinic/} and \url{http://www.shopizer.com/}.}, presented in Table~\ref{tab:targetsystems}, which summarizes the general characteristics of each target system. To prevent application bias in our results, we selected applications with different sizes (6.3--111.3 KLOC) and domains. Cloud Store, in particular, is developed mainly for performance testing and benchmarking, and follows the TPC-W performance benchmark standard. It is important to highlight that the DEV configuration was implemented by developers independently from the results of our previous work~\cite{Mertz2016} and, therefore, the \emph{Cacheability Pattern} was not considered in this implementation.

For Pet Clinic and Cloud Store, we used test cases written by their developers and developed test cases for Shopizer, for which they are unavailable. For the latter, we created test cases to cover searching, browsing, adding items to shopping carts, checking out, and editing products.

\begin{table}[t]
    \scriptsize
    \caption{Target systems of our study.}
    \label{tab:targetsystems}
    \centering
    \begin{tabular}{l p{2.8cm} rr p{4.4cm}}
      \toprule
      \textbf{Project} & \textbf{Domain} & \textbf{LOC} & \textbf{\# Files} & \textbf{Database Properties} \\
      \midrule
      Cloud Store & e-Commerce based on TPC-W benchmark & 7.6K & 98 & 300K customer data and 10K items \\ 
      Pet Clinic & Sample application & 6.3K & 72 & 6K vets, 10K owners and 13K pets \\ 
      Shopizer & e-Commerce & 111.3K & 946 & 300K customer data and 10K items \\ 
      \bottomrule
    \end{tabular}
\end{table}

We used Tomcat\footnote{\url{http://tomcat.apache.org/}} with 2G RAM dedicated its JVM as our web server and MySQL\footnote{\url{https://www.mysql.com/}} as the DBMS. We used two machines located within the same network, one machine for the DBMS and web server (16G RAM, Intel i7 2GHz), and one machine for JMeter\footnote{\url{http://jmeter.apache.org/}} (32G RAM, Intel i7 3.4GHz). The selected underlying caching framework is EhCache because it is used in all target applications. Moreover, we also used the same cache component configurations (TTL and in-memory cache size) as specified by developers in each application. Regarding the TTL implementation and its operation, we rely on the default behavior provided by the chosen cache provider.

\subsection{Results}

\subsubsection{RQ1. What is the performance improvement provided by our automated caching approach?} 

Based on our simulation, we observed that our approach (\textbf{AP}) improves the throughput of all target applications, in comparison with no use of caching (\textbf{NO}). Moreover, when compared with the application-level caching manually implemented by developers (\textbf{DEV}), our approach achieves at least similar performance. After a manual investigation, we concluded that our approach caches methods that lead to a higher number of hits than those chosen by developers. Consequently, caching them significantly reduced the network transfer time, and thus resulted in a substantial performance improvement. Thus, even if results were not as good as those obtained with DEV, they could be considered good because our approach automates an error-prone and time-consuming task performed by developers. Table~\ref{tab:results} shows the throughput obtained for each target application in simulations with five simultaneous users.

\begin{table}[t]
\scriptsize
\centering
\caption{Simulation Results: Throughput for Five Simultaneous Users.}
\label{tab:results}
\begin{tabular}{l rr rrr rr rrrr}
\toprule
& \multicolumn{1}{c}{\textbf{NO}} & \multicolumn{2}{c}{\textbf{DEV}} & \multicolumn{2}{c}{\textbf{AP}} \\ \cmidrule(l){2-2} \cmidrule(l){3-4} \cmidrule(l){5-6}
\textbf{Application} & \textbf{Throughput} & \multicolumn{2}{c}{\textbf{Throughput}} & \multicolumn{2}{c}{\textbf{Throughput}}\\ 
Cloud Store & 22.28 & 22.73 & (+2.01\%) & 22.90 & (+2.78\%)\\ 
Pet Clinic & 7.68 & 8.47 & (+10.28\%) & 9.01 & (+17.18\%)\\ 
Shopizer & 15.92 & 16.25 & (+2.07\%) & 16.96 & (+6.53\%)\\ 
\bottomrule
\end{tabular}
\end{table}

\begin{figure}[t]
\centering
\includegraphics[width=0.8\linewidth]{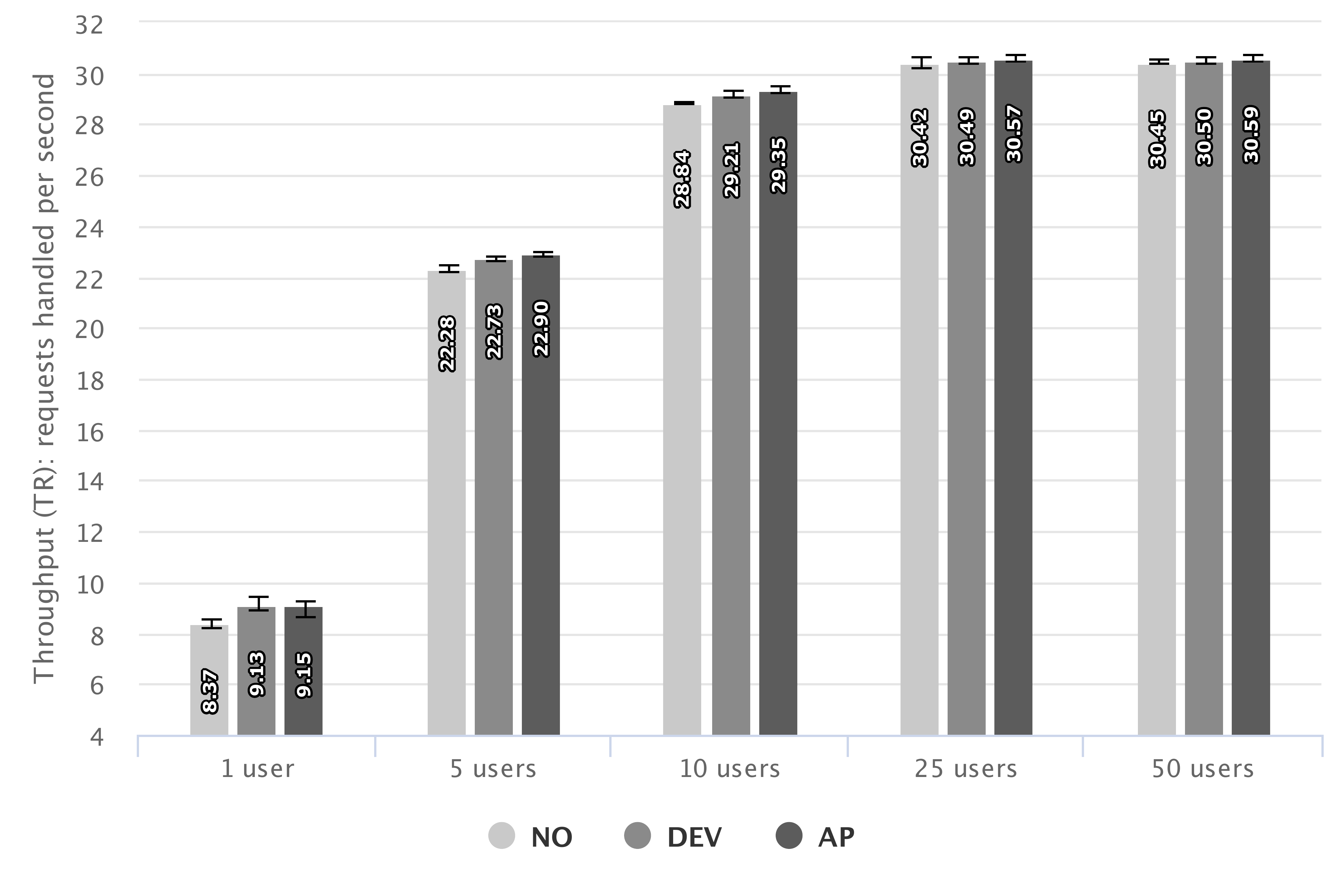}
\caption{Throughput by Caching Approach for Cloud Store.}
\label{fig:results1}
\end{figure}

\begin{figure}[t]
\centering
\includegraphics[width=0.8\linewidth]{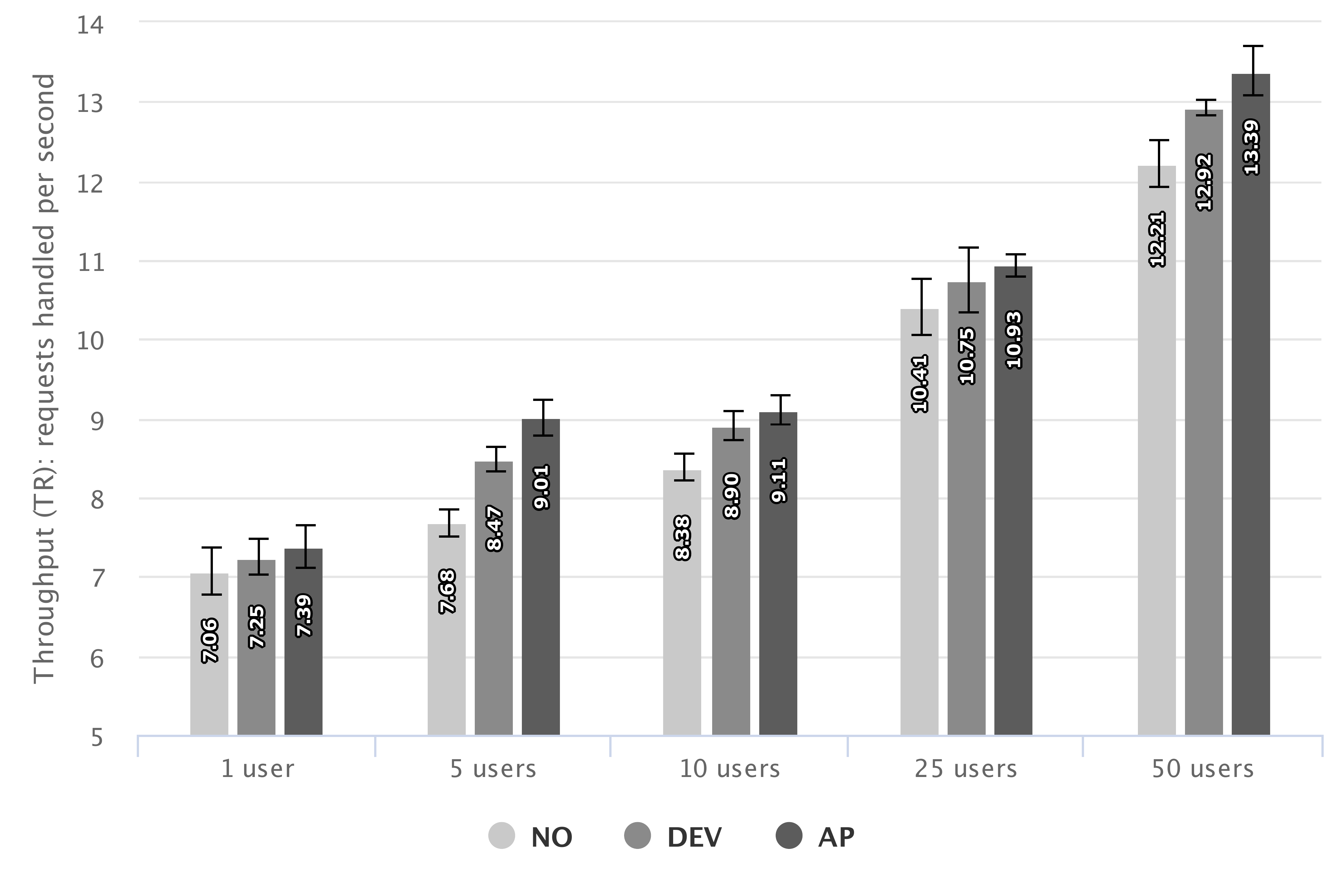}
\caption{Throughput by Caching Approach for Pet Clinic.}
\label{fig:results2}
\end{figure}

\begin{figure}[t]
\centering
\includegraphics[width=0.8\linewidth]{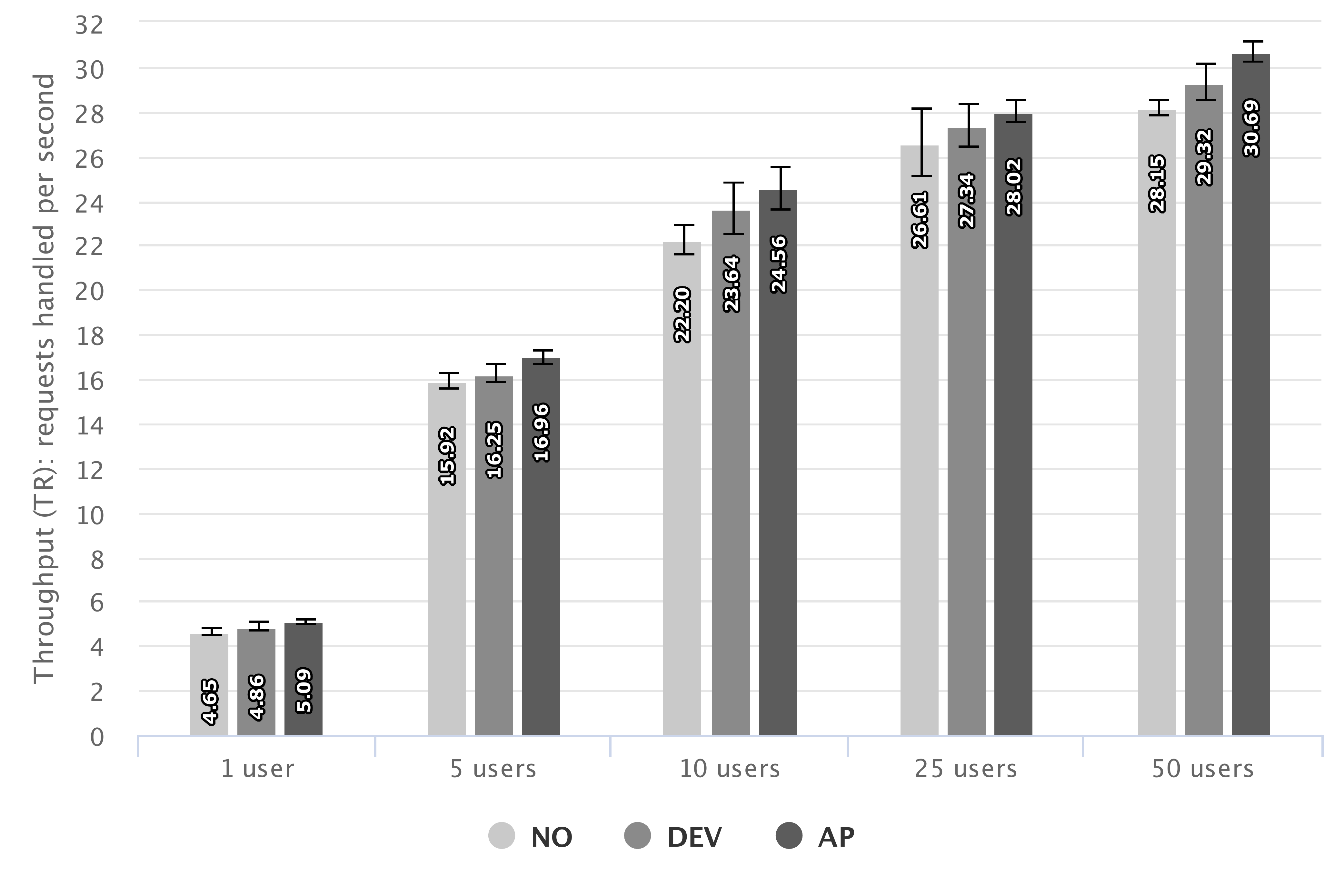}
\caption{Throughput by Caching Approach for Shopizer.}
\label{fig:results3}
\end{figure}

As conclusion, AP provides higher performance improvements (2.78\%--17.18\%), when compared to the improvements provided by developers with manual caching implementation. Figures~\ref{fig:results1}, \ref{fig:results2} and \ref{fig:results3} further complement the results with the throughput obtained with different amount of simultaneous users for each application. As can be seen, for the three studied applications, the throughput achieved by AP is higher or at least similar to all DEV executions even considering different levels of stress on the application server. However, in general, only a few methods are usually cached through application-level techniques. Thus, the advantage of caching is usually limited in scope, but yet beneficial to the overall performance of the system.

Regarding different cache policies, the evaluation with LRU and LFU resulted in similar throughput and hits as our standard configuration, for all the applications. This is explained by the fact that in all executions of our approach no eviction was necessary because the number of cacheable items never reached the maximum cache size before the TTL of existing content expired (which has a value specified by developers). Thus, we limit ourselves to present only the results achieved by our standard configuration, i.e.\ a TTL-based eviction. The complete results are available elsewhere\footnote{\url{http://inf.ufrgs.br/prosoft/resources/2017/spe-aplcache-framework}}.

Table~\ref{tab:results2} presents the hit ratio and the total number of hits for each target application in simulations with five simultaneous users. When considering these metrics, DEV is used as the baseline for AP, and we observed that AP achieves good results in comparison with DEV. First, our approach shows a high hit ratio improvement for Cloud Store and Shopizer. This improvement is related to caching \emph{search} operations because developers cache all search combinations, and our approach only caches searches that can potentially improve application performance (according to our evaluation criteria). For Pet Clinic, AP achieves a lower hit ratio than DEV. However, the total number of hits is 43.14\% higher. Despite the decrease in hit ratio, it means that AP identified cacheable opportunities that in general provides more hits than misses, and thus it indicates that AP identifies good cacheable opportunities.

\begin{table}[t]
\scriptsize
\centering
\caption{Simulation Results: Hit Ratio and Total Number of Hits for Five Simultaneous Users.}
\label{tab:results2}
\begin{tabular}{l rr rr rr}
\toprule
& \multicolumn{2}{c}{\textbf{DEV}} & \multicolumn{4}{c}{\textbf{AP}} \\ \cmidrule(l){2-3} \cmidrule(l){4-7}
\textbf{Application} & \textbf{Hit Ratio} & \textbf{Total Hits} & \multicolumn{2}{c}{\textbf{Hit Ratio}} & \multicolumn{2}{c}{\textbf{Total Hits}}\\ 
Cloud Store & 74.25\% & 1179 & 97.42\% & (+31.20\%) & 2416 & (+104.90\%)\\ 
Pet Clinic & 99.58\% & 1122 & 92.70\% & (-6.90\%) & 1606 & (+43.14\%)\\ 
Shopizer & 73.44\% & 8991 & 99.98\% & (+36.13\%) & 35259 & (+292.11\%)\\ 
\bottomrule
\end{tabular}
\end{table}

As conclusion, our approach was able to discover cacheable method calls at runtime, based on the application workload, with an improvement of our baseline. Therefore, it can relieve the burden from developers of identifying and implementing caching. Furthermore, if there is no enough trust to allow our approach to automatically manage the cache in a production environment, its caching decisions can be used as a guideline to developers while developing application-level caching, which is not trivial mainly in large applications.

Our approach takes about 1 minute to analyze 2 million application traces in the machines with the described configuration. It is only required to derive the set of cacheable methods. Such process is executed in background and thus has minimal impact on the application. Note that it can alternatively be configured to run on a separate machine. Monitoring all method calls can also minimize the performance improvement of our approach. However, this depends on when and how traces are collected from the application. For example, samples can be collected during off-peak periods to not compromise the overall application performance.

\subsubsection{RQ2. What are the similarities and differences between decisions made by our automated approach and by developers?}

The results of our simulation showed that our approach can improve the performance of web applications. However, it is interesting to understand the causes of this improvement. Therefore, we now compare the number of caching opportunities that were selected and managed by our approach with the choice made by developers, implemented in the target applications. By making this comparison, we observed that our approach not only caches the methods selected by developers but many others, as shown in Table~\ref{tab:cached-methods}. However, the number of selected methods to be cached is small in comparison with all possible methods, as shown in the column AP.

\begin{table}
\footnotesize
\centering
\caption{Number of Cacheable Methods: Our Approach \emph{vs.} Human-made Decisions.}
\label{tab:cached-methods}
\begin{tabular}{lrrrrrr}
\toprule
\textbf{Application} & \textbf{Total} & \textbf{DEV} & \textbf{AP} & \textbf{Intersection} & \textbf{DEV Only} & \textbf{AP Only} \\ \midrule
Cloud Store & 812 & 4 & 8 & 4 & 0 & 4 \\ 
Pet Clinic & 205 & 1 & 4 & 1 & 0 & 3 \\ 
Shopizer & 5212 & 15 & 22 & 15 & 0 & 7 \\
\bottomrule
\end{tabular}
\end{table}

Results indicate that developers may be conservative while identifying cacheable methods and select only those that lead to a strong confidence that caching them increases will result in cache hits. For instance, in Pet Clinic the number of veterinaries is often the same, and thus it is the only cacheable method identified by developers. Such opportunities are always detected by our approach. However, our approach was able to identify more cacheable opportunities, which justifies the performance improvement. As conclusion, our approach identifies a higher number of cacheable methods (46.66\%--300\%), when compared to the total number of cached methods identified by developers. For large applications like Shopizer, manually analyzing and identifying possible cacheable methods may be time-consuming or even infeasible in practice.

Furthermore, while implementing a cache solution, developers usually select \emph{cacheable methods}, and thus any call to a cacheable method is cached. This can lead to a higher and inefficient use of the cache size because not all method calls are frequently called or expensive. Our approach, in contrast, can deal with specific method calls, leading to an optimal utilization of the cache infrastructure. Moreover, this makes the cache effectiveness less dependent on the cache replacement policy, because fewer method calls are added to the cache and, with adequate TTL and size, less eviction is needed to free space in the cache.

\subsection{Threats to Validity}

We now analyze the threats to the validity of our empirical evaluation. First, the performance benefits of caching highly depend on workloads. Consequently, it is possible that the adopted workload from the performance tests may not be representative enough, given that we do not make any assumption regarding the workload when conducting our experiments and rely on the randomness added to the tests. Nevertheless, our approach does not depend on a particular workload and can find cacheable methods with any pre-specified workload, which may evolve over time in real world scenarios. Therefore, even if the workload changes substantially and initial cacheable methods are no longer useful, our approach can adapt itself, automatically discarding old caching configurations and discovering a new set of cacheable methods. Second, our evaluation involves only three target applications. Therefore, results may not be generalizable. To address this threat, we selected open-source applications, with different sizes and domains, implemented by different developers.

\section{Limitations}
\label{sec:limitations}

Providing a caching solution requires dealing with many challenges other than that addressed in this paper, such as consistency management, replacement policies, and distributed infrastructures. Therefore, some challenges are out of the scope. Moreover, considering our challenge of deciding what to cache, we are aware of shortcomings of our approach. We next discuss these limitations of our work, and how they can be addressed.

Given that we monitor method inputs and outputs to make caching decisions, we assume that the output depends only on the provided inputs, i.e.\ the output is a function of the input. For example, an invoked method may not only provide an output but also change other objects, i.e.\ the state of the application. Therefore, if we cache this method, it will not be invoked but its result will be obtained from the cache, and these side effects of the method call will not be achieved, possibly leading the application to an inconsistent state.
In this hidden state cases, developers could annotate the code to guide the approach towards avoiding tracking and caching such methods. It is reasonable to assume that the identification of methods that cannot be cached is easier than those that can be cached to improve performance. A static analysis of the source code may be enough to detect such methods.

Many popular caching frameworks adopt a weak consistency approach as default and use a time-based expiration policy to invalidate cached methods. Such approach favors performance~\cite{Sun2016} and is easier than defining a hard-to-maintain but more robust invalidation process. Our approach does not cover the challenge of when to expire cached content in the cache, so our implementation currently provides weak consistency, but this should be customized by developers if a more robust policy is required.

The overhead of the data tracking activity was not the focus of our evaluation. However, its impact did not prevent our approach to deliver cacheable opportunities in a timely fashion. However, if this activity has an impact on the application execution that is unacceptable, it can be configured to collect only samples or be enabled only at specific times. For example, after the analysis and identification of cacheable methods, the monitoring can remain disabled until the performance decreases, which means that the previously identified opportunities are not useful anymore.

Other caching issues not addressed in this paper, such as concurrency, scheduling, and replacement policies, are part of our future work. Given that our goal is to identify and cache content, we rely on the underlying caching frameworks for solving these issues. We use default configurations of caching providers and provide means for developers to customize such configurations.

\section{Related Work}
\label{sec:related_work}


In this section, we present work focused on supporting the implementation of a caching solution, which is one of our challenges, followed by automated and adaptive approaches to assess the cacheability level of data at the application-level.

\subsection{Cache Implementation}

A research challenge in application-level caching is how to reduce the burden of implementing application-level caching from developers. As a consequence, implementation-focused approaches have been extensively proposed. These solutions can raise the abstraction level of caching and reduce a significant amount of cache-related code to be added to the base code.

These approaches take the form of supporting libraries and frameworks, providing useful and ready-to-use cache-related features. However, such solutions require code changes and a manual integration with web applications to exploit caching benefits i.e.\ developers are responsible for managing the cache. Examples of such solutions are distributed cache systems, e.g.\ Redis\footnote{\url{https://redis.io/}} and Memcached, and libraries that cache content locally, e.g.\ Spring Caching\footnote{\url{https://docs.spring.io/spring/docs/current/spring-framework-reference/html/cache.html}}, EhCache, Infinispan\footnote{\url{http://infinispan.org/}}, Caffeine and Rails low-level caching\footnote{\url{http://edgeguides.rubyonrails.org/caching\_with\_rails.html\#low-level-caching}}. In addition, as caching is essentially a cross-cutting concern~\cite{Bouchenak2006}, aspect-oriented programming (AOP)~\cite{Kiczales1997} have been explored in order to provide a flexible and easy-to-use solution, such as Jcabi-aspects\footnote{\url{http://aspects.jcabi.com/annotation-cacheable.html}}.

The drawback of traditional supporting libraries and frameworks is partially addressed by approaches that provide developers with ways to declare knowledge associated with the semantics of application code and data through annotations in the code, and then specific caching tasks are automated. Such solutions have a lower impact on the application, as it does not require to introduce code interleaved with its base code. For example, \emph{CacheGenie}~\cite{Gupta2011} and TxCache~\cite{Ports2010} provide cache abstractions based on a specific and simple declarative programming model where developers can indicate the methods that should be cached. Then the proposed approaches can automatically cache and invalidate the results. Similarly, Huang et al.~\cite{Huang2010} proposed a browser-side caching framework that allows developers to customize their caching strategies, such as expiration times, cache granularity, and replacement policies.

Totally seamless and transparent solutions are usually coupled to the application as a surrounding layer, such as database~\cite{Ravi2009} or proxy-level~\cite{Ali2011} caching approaches. In this context, \emph{EasyCache}~\cite{Wang2014} is a hybrid solution that combines properties of database caching and application-level caching to provide transparent cache pre-loading, access and consistency maintenance without extensive modifications to the application or a complete redesign of the database. Similarly, \emph{AutoWebCache}~\cite{Bouchenak2006} can also be seen as a hybrid approach that associates the back-end databases and the dynamic web pages at the front-end. AutoWebCache uses AOP with pre-defined caching aspects to add caching of dynamic web pages to a servlet-based web application that interfaces a database with JDBC, managing consistency between such components through effective cache invalidation policies.

Although implementation-centered approaches can raise the abstraction level of caching, they still demand cache reasoning on developers, such as deciding whether to cache content.

\subsection{Identification of Caching Opportunities}

As mentioned before, application-level caches allow arbitrary content to be cached, and opportunities for caching thus emerge in the most diverse parts of the application. A popular approach to support developers while admitting content to the cache is to recommend caching opportunities. In this context, application profiling is usually the technique adopted, such as in \emph{MemoizeIt}~\cite{DellaToffola2015}, which compares inputs and outputs of method calls and gives a set of redundant operations. To avoid comparing objects in detail, MemoizeIt first compares objects without following any object references, and then iteratively increases the depth of exploration while shrinking the set of considered methods. Also by analyzing method calls, Maplesden et al.~\cite{Maplesden2015} proposes an approach that analyzes the smallest parent distance among common parents of a method to identify repeated patterns, which are named \emph{subsuming methods}.

Xu~\cite{Xu2012} addresses the problem of frequent creation of data structures, whose the lifetimes are not connected, but the content is always the same. Then, a list of top allocation sites that create such data structures are reported. Similarly, \emph{Cachetor}~\cite{Nguyen2013} addresses repeatedly computations by providing a runtime profiling tool that uses a combination of dynamic dependency and value profiling to suggest spots of invariant data values.

Although these approaches can potentially relieve reasoning burden from developers, such recommendations should be analyzed and implemented by them, integrating the appropriate cache logic into the application. Thus, approaches that identify but also automate the caching of cacheable content have been proposed. However, besides the analysis of the application behavior, such approaches also employ mechanisms to manage cache and application at runtime.

In this context, \emph{IncPy}~\cite{Guo2011} achieve this by implementing a technique as a custom open-source Python interpreter. IncPy explores the repetitive creation and processing of intermediate data files, which should be properly managed by developers to multiple dependencies between their code and data files. Otherwise, their analyses produce wrong results. To enable developers to iterate quickly without needing to manage intermediate data files, they added a set of dynamic analyses to the programming language interpreter. IncPy then automatically caches the results of long-running pure method calls to disk, manages dependencies between code and on-disk data, and later reuses results, rather than re-executing those methods. Furthermore, this approach also allows developers to customize the execution by inserting annotations, which can force IncPy to always or never cache particular methods. Our framework is similar in this sense because it also provides annotations to developers aggregate domain-specific knowledge.

Another approach that deals with the same problem is \emph{CacheOptimizer}~\cite{Chen2016}, which is based on two analysis parts. First, it performs a static code analysis to identify possible caching spots in database queries. Then, CacheOptimizer monitors readable weblogs and creates mappings between current workload and database access to decide which and when database access is cacheable. Although this approach addresses method caching opportunities, it is focused on database-centric web applications; thus, only database-related methods are supported. Our approach share commonalities with CacheOptimizer; however, we focus on general application methods while searching for cacheable options.

Baeza-Yates et al.~\cite{Baeza-Yates2007} addressed the identification of cacheable content in a different way, by filtering infrequent queries of web search engines, which cause a reduction in hit ratio because caching them often does not lead to hits. Therefore, this approach can prevent infrequent queries from taking space of more frequent queries in the cache. The proposed cache monitors stateless and stateful information of query executions and has two fully-dynamic parts. The first part is an admission controlled cache that only admits queries that the admission policy classifies as future cache hits. All queries that the admission policy rejects are admitted to the second part of the cache, an uncontrolled cache. Both caches implement a regular cache policy, more specifically, LRU. The uncontrolled cache can, therefore, manage queries that are infrequent but appear in short bursts, considering that the admission policy will reject queries that it concludes to be infrequent. Thus, the uncontrolled cache can handle cases in which infrequent queries may be asked again by the same user or within a short period. It guarantees that fewer infrequent queries enter the controlled cache, which is expected to handle temporal locality better.

Also focusing on filtering content, \emph{TinyLFU}~\cite{Einziger2017} uses an approximate LFU structure, which maintains a representation of the access frequency of recently accessed contents, to boost the admission effectiveness of caches. TinyLFU acts reactively when the cache is full and decides whether it is worthwhile admitting content, considering the cost of an eviction and the usefulness of the new content. This approach is currently available to developers as a replacement policy of the Caffeine caching framework.

Sajeev and Sebastian~\cite{Sajeev2010b} proposed a semi-intelligent admission technique using the multinomial logistic regression (MLR) classifier. They used previously collected traces to build an MLR model, which can classify the content worthiness class. Such class is achieved by computing six different parameters, which refers to the traffic and the content properties. Then, at runtime, every incoming content to the cache has its worthiness class computed or updated, and it is used in an admission control mechanism, based on thresholds, to decide whether the content should be admitted.

\subsection{Discussion}


Different from programmatic solutions to application-level caching implementation, such as traditional caching libraries and frameworks, our proposed caching approach does not require any additional implementation, detaching caching concerns from the application. Furthermore, our framework can automatically capture all the application-specific information needed to achieve its objectives (i.e.\ select and cache cacheable content), as opposed to other solutions~\cite{Ports2010,Gupta2011}, which demand input and configuration from developers. Despite not being required, our framework also allows developers to provide additional knowledge by using a declarative approach, which can improve the solution.

In addition to the traditionally explored access history and cache-related statistics in existing admission solutions~\cite{Baeza-Yates2007,Sajeev2010b,Einziger2017}, such as frequency and recency, our approach also considers caching metadata retrieved from the application during its execution, such as cost to retrieve, user sessions, cache size and data size. Such metadata are in fact information that developers use while designing and implementing application-level caching, and thus enrich the application model with valuable application-specific information regarding the applicability of caching. Our results provide evidence of the value of this kind of information.

Apart from exploring new kinds of metadata, our approach also differs from these admission solutions in the granularity of cached content. We address complex logic and personalized content, which are produced and handled by method calls of applications, as opposed to whole web pages or database queries. Moreover, we consider all method calls as cacheable options and do not focus on specific methods or types of web applications, as some approaches do~\cite{Guo2011,Maplesden2015,DellaToffola2015,Chen2016}.

\section{Conclusion}
\label{sec:conclusion}

Web developers often make use of different application-level caching frameworks to improve performance. However, they need to reason about what to cache and continually revise caching decisions. Otherwise, the designed caching may not achieve performance requirements. In this paper, we proposed an automated approach that manages the cache according to data collected at runtime. We also presented a seamless framework that implements our approach and detaches caching concerns from the application. Our approach combines a monitoring and analysis of system execution, and a runtime control loop to deal with caching concerns. As a result, we can provide an application-specific solution without pre-defined thresholds or assumptions.

We evaluated our approach with three open source applications, and results indicate it may improve their throughput up to 17.18\%, in comparison with the caching configuration designed by developers. Alternatively, our approach can be used as a supporting tool to help developers select cacheable content, given that developing a caching solution may require extensive and scattered code changes, which can be an error-prone and time-consuming task for developers. Although our approach was implemented specifically for Java-based web applications, it is generic enough to be used with any programming language.

Future work involves extending the approach to deal with other caching issues towards reducing the developers' effort when designing and implementing application-level caching. Although the processing phase of our approach seems to be fast enough to provide cacheable opportunities in a timely fashion, the overhead of the data tracking activity should be further evaluated regarding scalability. Therefore, techniques for reducing the amount of data required to identify cacheable methods should be investigated, towards reducing the overhead and providing a faster model building. Finally, we intend to study the execution of developer-written tests, such as unit and acceptance tests to find a less confident but yet representative list of cacheable methods. Thus, the application can benefit from caching since its first releases.





\ack We thank the anonymous reviewer who were asked by Software: Practice and Experience to review this article, as well as the editor. They provided constructive feedback that was used to expand our research and improve this article extensively. Jhonny Mertz would like to thank CNPq grant 131550/2015-2. Ingrid Nunes would like to thank for research grants CNPq ref. 303232/2015-3, CAPES ref. 7619-15-4, and Alexander von Humboldt, ref. BRA 1184533 HFSTCAPES-P.

\bibliographystyle{wileyj}
\bibliography{main}

\begin{thebibliography}{10}
\providecommand{\url}[1]{\texttt{#1}}
\providecommand{\urlprefix}{URL }
\expandafter\ifx\csname urlstyle\endcsname\relax
  \providecommand{\doi}[1]{doi:\discretionary{}{}{}#1}\else
  \providecommand{\doi}{doi:\discretionary{}{}{}\begingroup
  \urlstyle{rm}\Url}\fi

\bibitem{Ports2010}
Ports DRK, Clements AT, Zhang I, Madden S, Liskov B. {Transactional Consistency
  and Automatic Management in an Application Data Cache}. \emph{Proceedings of
  the 9th USENIX Symposium on Operating Systems Design and Implementation},
  USENIX Association: CA, USA, 2010; 279--292.

\bibitem{Chen2016}
Chen TH, Shang W, Hassan AE, Nasser M, Flora P. {CacheOptimizer: Helping
  Developers Configure Caching Frameworks for Hibernate-based Database-centric
  Web Applications}. \emph{Proceedings of the 2016 24th ACM SIGSOFT
  International Symposium on Foundations of Software Engineering - FSE 2016},
  ACM Press: New York, New York, USA, 2016; 666--677,
  \doi{10.1145/2950290.2950303}.

\bibitem{Radhakrishnan2004}
Radhakrishnan G. {Adaptive application caching}. \emph{Bell Labs Technical
  Journal}  may 2004; \textbf{9}(1):165--175, \doi{10.1002/bltj.20011}.

\bibitem{Negrao2015}
Negr{\~{a}}o AP, Roque C, Ferreira P, Veiga L. {An adaptive semantics-aware
  replacement algorithm for web caching}. \emph{Journal of Internet Services
  and Applications}  feb 2015; \textbf{6}(1):4,
  \doi{10.1186/s13174-015-0018-4}.

\bibitem{DellaToffola2015}
{Della Toffola} L, Pradel M, Gross TR. {Performance problems you can fix: a
  dynamic analysis of memoization opportunities}. \emph{Proceedings of the 2015
  ACM SIGPLAN International Conference on Object-Oriented Programming, Systems,
  Languages, and Applications - OOPSLA 2015}, ACM Press: New York, New York,
  USA, 2015; 607--622, \doi{10.1145/2814270.2814290}.

\bibitem{Labrinidis2009}
Labrinidis A. {Caching and Materialization for Web Databases}.
  \emph{Foundations and Trends{\textregistered} in Databases}  mar 2009;
  \textbf{2}(3):169--266, \doi{10.1561/1900000005}.

\bibitem{Ravi2009}
Ravi J, Yu Z, Shi W. {A survey on dynamic Web content generation and delivery
  techniques}. \emph{Journal of Network and Computer Applications}  sep 2009;
  \textbf{32}(5):943--960, \doi{10.1016/j.jnca.2009.03.005}.

\bibitem{Mertz2017a}
Mertz J, Nunes I. {Understanding Application-level Caching in Web Applications:
  a Comprehensive Introduction and Survey of State-of-the-art Approaches}.
  \emph{ACM Computing Surveys (CSUR)}  2017; \textbf{50}(6):34,
  \doi{10.1145/3145813}.

\bibitem{Li2006a}
Li L, Niu C, Zheng H, Wei J. {An adaptive caching mechanism for Web services}.
  \emph{Proceedings of the International Conference on Quality Software}, IEEE:
  Beijing, China, 2006; 303--310, \doi{10.1109/QSIC.2006.9}.

\bibitem{Guerrero2011}
Guerrero C, Juiz C, Puigjaner R. {Improving web cache performance via adaptive
  content fragmentation design}. \emph{Proceedings of the IEEE International
  Symposium on Network Computing and Applications}, IEEE: Cambridge, USA, 2011;
  310--313, \doi{10.1109/NCA.2011.55}.

\bibitem{Amza2005}
Amza C, Soundararajan G, Cecchet E. {Transparent caching with strong
  consistency in dynamic content web sites}. \emph{Proceedings of the 19th
  annual international conference on Supercomputing}, ACM Press: New York, New
  York, USA, 2005; 264, \doi{10.1145/1088149.1088185}.

\bibitem{Baeza-Yates2007}
Baeza-Yates R, Junqueira F, Plachouras V, Witschel H. {Admission Policies for
  Caches of Search Engine Results}. \emph{Proceedings of the 14th international
  conference on String processing and information retrieval}, vol. 4726,
  Springer-Verlag: Santiago, Chile, 2007; 74--85,
  \doi{10.1007/978-3-540-75530-2_7}.

\bibitem{Mertz2016}
Mertz J, Nunes I. {A Qualitative Study of Application-Level Caching}.
  \emph{IEEE Transactions on Software Engineering}  sep 2017;
  \textbf{43}(9):798--816, \doi{10.1109/TSE.2016.2633992}.

\bibitem{Gupta2011}
Gupta P, Zeldovich N, Madden S. {A trigger-based middleware cache for ORMs}.
  \emph{Proceedings of the 12th ACM/IFIP/USENIX International Middleware
  Conference}, vol. 7049 LNCS, Springer Berlin Heidelberg: Lisbon, Portugal,
  2011; 329--349, \doi{10.1007/978-3-642-25821-3_17}.

\bibitem{Santhanakrishnan2006}
Santhanakrishnan G, Amer A, Chrysanthis PK. {Self-tuning caching: The universal
  caching algorithm}. \emph{Software - Practice and Experience}  sep 2006;
  \textbf{36}(11-12):1179--1188, \doi{10.1002/spe.755}.

\bibitem{Basili1994}
Basili VR, Caldiera G, Rombach HD. {The goal question metric approach}.
  \emph{Encyclopedia of Software Engineering}  1994; \textbf{2}:528--532,
  \doi{10.1.1.104.8626}.

\bibitem{Binder2000}
Binder R. \emph{{Testing object-oriented systems: models, patterns, and
  tools}}. Addison-Wesley, 2000.

\bibitem{Sun2016}
Sun H, Xiao B, Wang X, Liu X. {Adaptive trade-off between consistency and
  performance in data replication}. \emph{Software - Practice and Experience}
  nov 2016; \doi{10.1002/spe.2462}.

\bibitem{Bouchenak2006}
Bouchenak S, Cox A, Dropsho S, Mittal S, Zwaenepoel W. {Caching Dynamic Web
  Content: Designing and Analysing an Aspect-Oriented Solution}.
  \emph{Proceedings of the ACM/IFIP/USENIX International Conference on
  Middleware}, \emph{Lecture Notes in Computer Science}, vol. 4290, van Steen
  M, Henning M (eds.), Springer Berlin Heidelberg: Berlin, Heidelberg, 2006;
  1--21, \doi{10.1007/11925071}.

\bibitem{Kiczales1997}
Kiczales G, Lamping J, Mendhekar A, Maeda C, Lopes C, Loingtier JM, Irwin J.
  {Aspect-oriented programming}. \emph{European conference on object-oriented
  programming}  1997; \textbf{1241/1997}(June):220--242,
  \doi{10.1007/BFb0053381}.

\bibitem{Huang2010}
Huang J, Liu X, Zhao Q, Ma J, Huang G. {A browser-based framework for data
  cache in web-delivered service composition}. \emph{Proceedings of the IEEE
  International Conference on Service-Oriented Computing and Applications},
  IEEE: Perth, Australia, 2010; 1--8, \doi{10.1109/SOCA.2010.5707138}.

\bibitem{Ali2011}
Ali W, Shamsuddin SM, Ismail AS. {A survey of Web caching and prefetching}.
  \emph{International Journal of Advances in Soft Computing and Its
  Applications}  mar 2011; \textbf{3}(1):18--44.

\bibitem{Wang2014}
Wang W, Liu Z, Jiang Y, Yuan X, Wei J. {EasyCache: a transparent in-memory data
  caching approach for internetware}. \emph{Proceedings of the 6th Asia-Pacific
  Symposium on Internetware on Internetware}, ACM Press: New York, New York,
  USA, 2014; 35--44, \doi{10.1145/2677832.2677837}.

\bibitem{Maplesden2015}
Maplesden D, Tempero E, Hosking J, Grundy JC. {Subsuming Methods}.
  \emph{Proceedings of the 6th ACM/SPEC International Conference on Performance
  Engineering - ICPE '15}, ACM Press: New York, New York, USA, 2015; 175--186,
  \doi{10.1145/2668930.2688040}.

\bibitem{Xu2012}
Xu G. {Finding reusable data structures}. \emph{Proceedings of the ACM
  international conference on Object oriented programming systems languages and
  applications - OOPSLA '12}  nov 2012; \textbf{47}(10):1017,
  \doi{10.1145/2384616.2384690}.

\bibitem{Nguyen2013}
Nguyen K, Xu G. {Cachetor: detecting cacheable data to remove bloat}.
  \emph{Proceedings of the 9th Joint Meeting on Foundations of Software
  Engineering}, ACM Press: New York, New York, USA, 2013; 268,
  \doi{10.1145/2491411.2491416}.

\bibitem{Guo2011}
Guo PJ, Engler D. {Using Automatic Persistent Memoization to Facilitate Data
  Analysis Scripting}. \emph{Proceedings of the 2011 International Symposium on
  Software Testing and Analysis}, ACM Press: New York, New York, USA, 2011;
  287--297, \doi{10.1145/2001420.2001455}.

\bibitem{Einziger2017}
Einziger G, Friedman R, Manes B. {TinyLFU: A Highly Efficient Cache Admission
  Policy}. \emph{ACM Transactions on Storage}  nov 2017; \textbf{13}(4):1--31,
  \doi{10.1145/3149371}.

\bibitem{Sajeev2010b}
Sajeev GP, Sebastian MP. {Building a semi intelligent web cache with light
  weight machine learning}. \emph{Proceedings of the IEEE International
  Conference on Intelligent Systems}, IEEE: Xiamen, China, 2010; 420--425,
  \doi{10.1109/IS.2010.5548373}.

\end{thebibliography}
\end{document}